\documentclass[%
 aip,
 amsmath,amssymb,
 twocolumn,
 reprint,%
]{revtex4-2}

\usepackage{graphicx}
\usepackage{dcolumn}
\usepackage{bm}
\usepackage{xcolor}

\usepackage{booktabs}
\usepackage{makecell}
\usepackage{enumitem}

\usepackage[T1]{fontenc}
\usepackage{textcomp,gensymb} 

\usepackage{hyperref} 
\hypersetup{
    colorlinks=true,
    linkcolor=blue,
    filecolor=magenta,      
    urlcolor=cyan,
}
\begin{document}


\title{Characterization of the stability and dynamics of a laser-produced plasma expanding across strong magnetic field}

\author{Weipeng Yao} 
\email{yao.weipeng@polytechnique.edu}
\affiliation{LULI - CNRS, CEA, Sorbonne Universit\'e, Ecole Polytechnique, Institut Polytechnique de Paris - F-91128 Palaiseau cedex, France}
\affiliation{Sorbonne Universit\'e, Observatoire de Paris, Universit\'e PSL, CNRS, LERMA, F-75005, Paris, France}

\author{Julien Capitaine} 
\affiliation{Sorbonne Universit\'e, Observatoire de Paris, Universit\'e PSL, CNRS, LERMA, F-75005, Paris, France}

\author{Benjamin Khiar}
\affiliation{Office National d'Etudes et de Recherches A\'erospatiales (ONERA), Palaiseau 91123, France}

\author{Tommaso Vinci}
\affiliation{LULI - CNRS, CEA, Sorbonne Universit\'e, Ecole Polytechnique, Institut Polytechnique de Paris - F-91128 Palaiseau cedex, France}

\author{Konstantin Burdonov} 
\affiliation{LULI - CNRS, CEA, Sorbonne Universit\'e, Ecole Polytechnique, Institut Polytechnique de Paris - F-91128 Palaiseau cedex, France}
\affiliation{Sorbonne Universit\'e, Observatoire de Paris, Universit\'e PSL, CNRS, LERMA, F-75005, Paris, France}
\affiliation{IAP, Russian Academy of Sciences, 603950, Nizhny Novgorod,  Russia}

\author{J\'er\^ome B\'eard}
\affiliation{LNCMI, UPR 3228, CNRS-UGA-UPS-INSA, Toulouse 31400, France}

\author{Julien Fuchs} 
\affiliation{LULI - CNRS, CEA, Sorbonne Universit\'e, Ecole Polytechnique, Institut Polytechnique de Paris - F-91128 Palaiseau cedex, France}

\author{Andrea Ciardi} 
\affiliation{Sorbonne Universit\'e, Observatoire de Paris, Universit\'e PSL, CNRS, LERMA, F-75005, Paris, France}

\date{\today}

\begin{abstract}

Magnetized laser-produced plasmas are central to many new studies in laboratory astrophysics, inertial confinement fusion, and in industrial applications.
Here we present the results of large-scale, three-dimensional magneto-hydrodynamic simulations of the dynamics of a laser-produced plasma expanding into a transverse magnetic field with strength of tens of Tesla. The simulations show the plasma being confined by the strong magnetic field into a slender slab structured by the magnetized Rayleigh-Taylor instability that develops at the plasma-vacuum interface. We find that by perturbing the initial velocity of the plume the slab can develops kink-like motion which disrupt its propagation.

\end{abstract}

\maketitle

\section{\label{sec:intro}INTRODUCTION}

Plasma flow across magnetic fields exists ubiquitously around the Universe \citep{bernhardt1987observations}. The stability and dynamics of such plasma flows are of paramount importance in understanding the deceleration, trapping, and heating of the plasma in the magnetic field.

Thanks to the development of high-power lasers coupled to high-strength magnetic fields devices  \citep{fujioka2013kilotesla,albertazzi2013production}, we are now able to investigate the interaction between plasma and strong magnetic field in the laboratory in a controllable and well diagnosed environment \citep{revet2017laboratory, schaeffer2017high, kuramitsu2018magnetic}. In addition, in the field of inertial confinement fusion (ICF), there is increasing need for a detailed understanding of plasma dynamics in the presence of strong magnetic fields \citep{slutz2012high,davies2017laser,filippov2021enhanced}, e.g. for cross-magnetic-field transport processes \citep{froula2007quenching}.

Besides the strength of the applied magnetic field, its relative direction (with respect to the plasma flow) plays an important role in the stability and dynamics of these plasmas. For example, in an axisymmetric scenario (when the two directions are aligned with each other), the plasma expansion is collimated into a stable jet  \citep{ciardi2013astrophysics,albertazzi2014laboratory,higginson2017detailed}. However for the case of a magnetic field transverse to the plasma flow, both stable \citep{plechaty2013focusing} and unstable  \citep{mostovych1989laser} flows were reported in the literature, and there has not been a clear understanding and detailed characterization of this issue yet. In the following, we briefly review the former studies of the interaction between laser-produced plasma and transverse magnetic field, before detailing our present contribution.

Over the past few decades, much effort has been devoted to investigate the overall dynamics of laser-produced plasma expansion across a magnetic field. We note that as our understanding of the underlying physics has progressed over time, this has been accompanied by progress in experimental capabilities, namely the available laser energy and magnetic field strength. To our knowledge, the earliest laboratory investigation on this subject can be traced back to the early 1970s \citep{bruneteau1970experimental,matoba1971motion}, in which the laser energy was up to 2.4~J 
and the magnetic field strength was $\lesssim6$~T. In these experiments, both plasma confinement and flow across the magnetic field were observed. The flow was explained, from a kinetic point of view, as the $\mathbf{E} \times \mathbf{B}$ drift, in which the electric field $\mathbf{E}$ was the result of charge-separation at the front of the expanding plasma. Notably, no instabilities were observed.

When the laser energy increased to about 7~J, some ``wings'' were uncovered at the leading edge of the plasma and a number of ``ripples'' inside the plasma were observed \citep{jellison1981resonant}, which drew the attention to the study of instabilities.
At higher laser energies, $\sim30$~J, a flute-modes instability (i.e. modes with wave-vector perpendicular to the external magnetic field) was observed and analysed carefully \citep{okada1981behaviour,ripin1987large,ripin1993sub}. 
In addition, bifurcation or splitting of the flute tips was observed during the nonlinear phase of the instabilities.
It was also reported that the characteristic wavelength of the flute-like structure was approximately independent of the magnetic field strength \citep{mostovych1989laser}. These instabilities have been attributed to the lower hybrid drift instability (LHDI)\citep{winske1989development, huba1990stability} because of the large ion Larmor radius ($> 2$ cm) and the relatively low collisionality associated with the low plasma densities produced ($< 10^{14}$ cm$^{-3}$).
Furthermore, as the laser energy increased, besides the LHDI, other modes of plasma instability were proposed, namely, the electron-ion hybrid instability \citep{peyser1992electron}, the unmagnetized Rayleigh-Taylor instability \cite{hassam1990nonlinear,zakharov2006role}, and the large Larmor radius instability \citep{tang2020observation}.

Note that in the above pioneering works, the magnetic fields were no more than tens of Tesla for laser energies of tens of Joules. In fact, if the laser energy is increased further but the magnetic field strength is not, the plasma will not be confined within the characteristic spatial and temporal scales of the experiment any more\cite{ciardi2013astrophysics}. Thus, laser energies of few tens of Joule and magnetic field strength up to few tens of Tesla are the parameter range we focus on here. 
In addition, owing to the limited availability of diagnostics in these early works, the plasma density, ionization state, temperature, and local magnetic field strength could not be assessed precisely. Since these are necessary to accurately characterize the laser-produced plasma and the magnetic field environment, it was hard to pinpoint the precise mechanism (kinetic or fluid) to understand the plasma propagation across the magnetic field and its dynamics. A large increase in the magnetic field strength was realized by using pulsed power generators to achieve up to 17 T, but in these experiments no instabilities were observed, at least in the plane perpendicular to the magnetic field where the observations were made\cite{plechaty2010penetration}.


Recently, a detailed picture of the interaction between a laser-produced plasma flow and a strong transverse magnetic field was presented\citep{khiar2019laser}. The plasma plume was observed to be confined into a slender, rapidly elongating slab, and structured by the magnetized Rayleigh-Taylor instability (MRTI).


In spite these efforts a comprehensive study of the 3D dynamics of the plasma over long time and spatial scales is still lacking. Here we provide the first such characterization using large-scale 3D resistive-MHD simulation with the code GORGON \citep{chittenden2004x, ciardi2007evolution}. 

This paper is organized as follows: in Sec.~\ref{SIMU}, we present our numerical model and simulation setups;
then, in Sec.~\ref{demo}, we detail the overall evolution of the plasma plume and characterize the key parameters in detail; in Sec.~\ref{formation}, we investigate how the slab formation can be affected when applying stronger magnetic fields and asymmetric initial perturbations; in Sec.~\ref{propagation}, we analyse the possible reasons for the disruptions that affect the plasma slab propagation, provided a strong enough magnetic field is applied. Finally in Sec.~\ref{CONC.}, we give our conclusions.

\section{NUMERICAL SETUP}
\label{SIMU}

\begin{figure}[htp]
    \centering
    \includegraphics[width=0.48\textwidth]{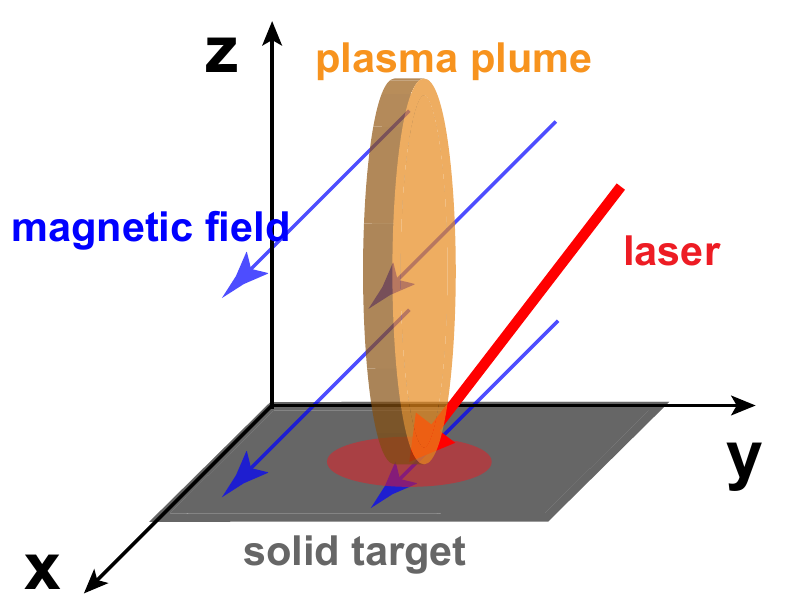}
    \caption{\textbf{Schematic diagram of the simulation setup.} The solid target is in the xy-plane, the externally applied magnetic field direction is along the x-axis, and the plasma plume expands along the z-axis.}
    \label{fig:setup}
\end{figure}

We simulate the typical laser conditions of experiments performed on ELFIE laser \cite{albertazzi2014laboratory,revet2017laboratory,khiar2019laser,filippov2021enhanced}. The schematic of the simulation setup is shown in Fig.\ref{fig:setup}, the solid C$_2$F$_4$ target (in grey) is irradiated with a laser pulse (in red) with 17~J energy, 0.5~ns pulse duration (at Full-Width-Half-Maximum), and 1.057~$\mu$m laser wavelength. The focal spot diameter is 750~$\mu$m, and the on-target intensity is about $7.7\times 10^{12}$~W/cm$^{2}$. The magnetic field (in blue) is uniformly applied along the x-direction with $B_x = 30$~T, and we will focus on the evolution of the plasma plume (in orange).

The simulation box is defined by a uniform Cartesian grid of dimension 8 mm  $\times$ 8 mm $\times$ 30 mm and the number-of-cells is equal to $400 \times 400 \times 1500 = 2.4\times 10^8$, the spatial resolution is $dx = dy = dz = 20$~$\mu$m, and the simulation duration is $50$ ns. Both the box size (along the $z$-direction) and the simulation duration are increased with respect to our former paper \citep{khiar2019laser} in order to investigate the late-time slab propagation. We here consider ``outflow'' boundary conditions. GORGON solves the resistive MHD equations, uses a vector potential formalism and retain the displacement current in vacuum. This allows the code to model a computational vacuum whose cut-off density is set to $10^{-4}$ kg/m$^{3}$.

The interaction between the laser and the solid target is performed using the DUED code \citep{atzeni2005fluid}, which solves the single-fluid, three-temperatures equations in two-dimensional axisymmetric, cylindrical geometry in Lagrangian form. The code uses the material properties of a two-temperatures equation of state (EOS) model including solid state effects, and a multi-group flux-limited radiation transport module with tabulated opacities. The laser-plasma interaction is performed in the geometric optics approximation including inverse-Bremsstrahlung absorption. At the end of the laser pulse (about 1~ns), the plasma profiles of density, momentum and temperature from the DUED simulations are remapped onto the 3D Cartesian grid of GORGON. The purpose of this hand-off is to take advantage of the capability of the Lagrangian code to achieve very high resolution in modeling the laser-target interaction. A similar configuration has been used in former works \citep{ciardi2013astrophysics, albertazzi2014laboratory, higginson2017detailed, higginson2017enhancement,khiar2019laser}.

To remove the axis-symmetry imposed by the DUED simulations when remapping onto the GORGON grid, we introduce a uniformly distributed random perturbation on the plasma velocity components with maximum amplitude $\pm5$\% of the initial value. In addition, to reproduce the kink-like perturbations seen in the experiment\cite{khiar2019laser,filippov2021enhanced}, we impose an asymmetric modulation on the initial plasma velocity between the right and left side of the slab; we also explore Bessel-like modulations, the details of which are given in Sec.~\ref{propagation}.

\section{3D SIMULATIONS OF PLASMA EVOLUTION}
\label{demo}

\begin{figure}[htp]
    \centering
    \includegraphics[width=0.48\textwidth]{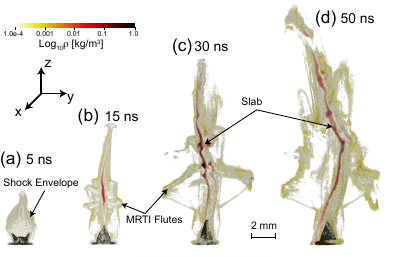}
    \caption{\textbf{Demonstration of the overall dynamics of the slab.} Panels (a), (b), (c) and (d) show the global 3D rendering of the decimal logarithm mass density at different times (i.e. {5, 15, 30 and 50 ns}) after the laser pulse.}
    \label{fig:3Ddemo}
\end{figure}

\begin{figure*}[htp]
    \centering
    \includegraphics[width=0.9\textwidth]{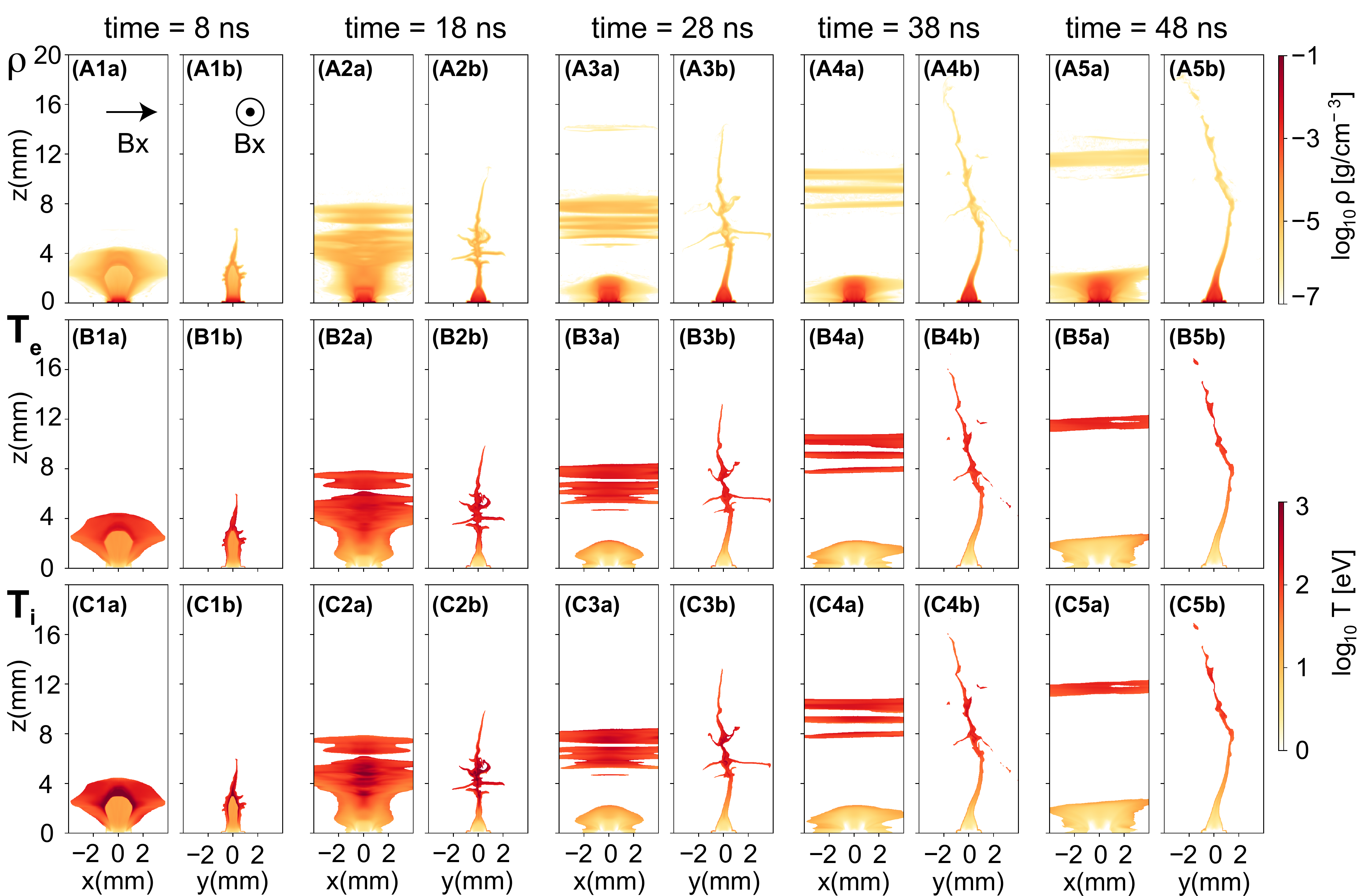}
    \caption{
    \textbf{The decimal logarithm of the mass density $\rho$ (A) and temperatures $T_e$ (B), $T_i$ (C) of the plasma plume sliced in the middle of the xz-plane (a) and the yz-plane (b) at varies times, i.e. 8/18/28/38/48 ns (1-5).}
    The temperatures share the same the color map in (B) and (C).
    The magnetic field directions are shown in (A1).
    }
    \label{fig:2Ddemo}
\end{figure*}

The global dynamics of the plasma plume over 50~ns is shown in Fig.~\ref{fig:3Ddemo}. 
The initial plasma plume expansion is shown in Fig.~\ref{fig:3Ddemo} (a), where we observe the typical diamagnetic cavity and the curved shock envelope that bounds the plasma \cite{khiar2019laser}; the development of the MRTI starts from Fig.~\ref{fig:3Ddemo} (b), where the cavity collapses along the y-direction and a slab is forming from the cavity tip; the development of the flutes and the extension of the slab along the z-direction can be seen in Fig.~\ref{fig:3Ddemo} (c), which already becomes kink-like unstable; and the slab propagation and bending are shown in Fig.~\ref{fig:3Ddemo} (d).  

Fig.~\ref{fig:2Ddemo} shows the 2D slices of the mass density $\rho$, electron temperature $T_e$, and ion temperature $T_i$ of the plasma plume in both the xz-plane and the yz-plane. Comparing the xz-plane with the yz-plane, it is clear that while the plasma plume has been confined in the yz-plane, it can freely expand in the xz-plane. The detailed plasma parameters are listed in Table~\ref{tab:paras}, together with the deduced dimensionless parameters. 

We now focus our attention on the collimation of the plasma plume into a slab and on its propagation.

\begin{table*}
\centering
\begin{tabular}{c@{\hskip 12\tabcolsep}c@{\hskip 8\tabcolsep}c@{\hskip 8\tabcolsep}c}
\toprule
\textbf{Time}                                            & \textbf{5 ns}          & \textbf{30 ns}               & \textbf{50 ns} \\
\textbf{Region}                                          & \textbf{Shock Envelope}  & \textbf{MRTI Flutes}        & \textbf{Slab}       \\ \midrule
\multicolumn{4}{c}{\textbf{Local Measurements}}                                                                                    \\ \midrule
\textbf{Characteristic length $\boldsymbol{L_0}$ {[}$\boldsymbol{\mu}$m{]}}            & 100                  & 100                   & 100                   \\
\textbf{Averaged Atomic Number A}                        & 17.3                  & 17.3                   & 17.3                   \\
\textbf{Effective Charge State $\boldsymbol{Z_{eff}}$}                         & 6.5                   & 8.0                    & 8.0                    \\
\textbf{Elec. Number Density $\boldsymbol{n_e}$ {[}cm$^{-3}${]}}      & $1.0 \times 10^{19}$                & $5.0 \times 10^{17}$                 & $1.0 \times 10^{18}$                 \\
\textbf{Ion Number Density $\boldsymbol{n_i}$ {[}cm$^{-3}${]}}        & $1.5 \times 10^{18}$                & $7.0 \times 10^{16}$                 & $1.5 \times 10^{17}$                 \\
\textbf{Mass Density $\boldsymbol{\rho}$ {[}g/cm$^{3}${]}}     & $4.0 \times 10^{-5}$                & $2.0 \times 10^{-6}$                 & $4.0 \times 10^{-6}$                 \\
\textbf{Elec. Temperature $\boldsymbol{T_e}$ {[}eV{]}}                & 150.0                 & 120.0                  & 100.0                  \\
\textbf{Ion Temperature $\boldsymbol{T_i}$ {[}eV{]}}                  & 400.0                  & 90.0                   & 100.0                  \\
\textbf{Flow Velocity $\boldsymbol{V_0}$ {[}km/s{]}}                  & 350.0                 & 150.0                  & 400.0                  \\
\textbf{Magnetic Field Strength B {[}T{]}}               & 25.0                  & 29.0                   & 30.0                   \\ \midrule
\multicolumn{4}{c}{\textbf{Calculated Dimensionless Parameters}}                                                                   \\ \midrule
\textbf{Elec. Collisionality $\boldsymbol{\lambda_{mfp,e} / L_0}$}    & 0.06                  & 0.6                   & 0.2                   \\
\textbf{Ion Collisionality $\boldsymbol{\lambda_{mfp,i} / L_0}$}      & 0.1                  & 0.1                   & 0.1                   \\
\textbf{Elec. Magnetization ($\boldsymbol{\lambda_{mfp,e}/r_{L,e}}$)} & 5.0                 & 680.0                 & 280.0                 \\
\textbf{Ion Magnetization ($\boldsymbol{\lambda_{mfp,i}/r_{L,i}}$)}   & 0.03                  & 0.5                   & 0.3                   \\
\textbf{Mach Number $\boldsymbol{M}$}                                 & 3.1                  & 1.6                   & 4.7                   \\
\textbf{Alfv\'enic Mach Number $\boldsymbol{M_A}$}                    & 3.3                 & 0.3                   & 1.0                   \\
\textbf{Magnetosonic Mach Number $M_{ms}$}                            & 2.3                  & 0.3                   & 1.0                   \\
\textbf{Plasma Thermal Beta $\boldsymbol{\beta_{ther}}$}              & 1.4                  & $3.0 \times 10^{-2}$                   & $5.0 \times 10^{-2}$                   \\
\textbf{Plasma Dynamic Beta $\boldsymbol{\beta_{dyn}}$}               & 22.0                & 0.1                   & 2.0                   \\ \bottomrule
\end{tabular}
\caption{\textbf{Measured plasma conditions and calculated dimensionless parameters} for the case with initial magnetic field $B_{x0} = 30$ T in different regions as indicated in Fig.~\ref{fig:3Ddemo} , i.e. the shock envelope at $t = 5$ ns, the MRTI flutes at $t = 30$ ns, and the slab at $t = 50$ ns. 
$\lambda_{mfp,s}$ ($s=i,e$ for ions and electrons, respectively) is the mean-free-path \cite{trubnikov1965particle}, $r_{L,s}$ is the Larmor radius. 
The Mach number is the ratio of the flow velocity over the sound velocity, while the Alfv\'{e}nic Mach Number is the ratio of the flow velocity over the Alfv\'{e}nic velocity.
The thermal (resp. dynamic) beta parameter is the ratio of the plasma thermal (resp. ram) pressure over the magnetic pressure. 
}
\label{tab:paras}
\end{table*}

\subsection{SLAB FORMATION}
\label{formation}

\begin{figure}[htp]
    \centering
    \includegraphics[width=0.42\textwidth]{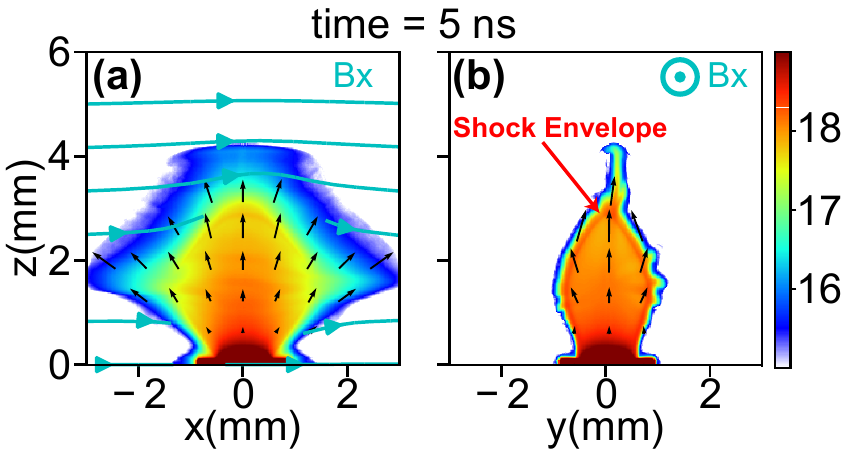}
    \caption{
    \textbf{Plasma expansion and diamagnetic cavity formation.} The decimal logarithm of the electron number density integrated along or perpendicular to the magnetic field, i.e. (a) the xz-plane and (b) the yz-plane at 5 ns after the laser ablation. 
    The color map corresponds to $\log_{10} \int n_e dx$ in cm$^{-2}$. 
    Black arrows show the direction and magnitude of the velocities, and
    light blue lines and arrows show the direction of the magnetic field in the xz-plane (they do not appear as continuous lines because they are taken out of the sliced plane).
    }
    \label{fig:Early_time_evo}
\end{figure}

In this section, we review and extend our former work \citep{khiar2019laser} to a higher magnetic field of 30 T, and an increased perturbation amplitude of the initial momentum on the right-half side of the target. 
As is quantified in Table~\ref{tab:paras}, the initial plasma expansion is dominated by the ram pressure, until almost stagnation around 5 ns, the plasma dynamic $\beta_{dyn} = \rho v^2 / (B^2/(2\mu_0))  \sim 22.0$, where $v$ is the fluid velocity, $B$ is the magnetic field strength and $\mu_0$ is the permeability in vacuum. This would result in an almost free expansion of the plasma in the xz-plane, as is detailed in Fig.~\ref{fig:Early_time_evo} (a).
The situation is quite different in the yz-plane, as is shown in Fig.\ref{fig:Early_time_evo} (b). There, the highly conductive plasma plume expands and pushes the magnetic field away. A pressure balance between the ram pressure of the plasma and the ambient magnetic pressure is achieved and this leads to the end of the diamagnetic cavity expansion \citep{ciardi2013astrophysics}. Because the plasma has velocities greater than the fast magneto-acoustic velocity, its deceleration leads to the formation of a curved shock envelope. The magnetosonic Mach number is $M_{ms} = V_0 / \sqrt{c_s^2 + v_A^2} \sim 2.3$, corresponding to a supersonic and super-Alfv\'enic regime (where $c_s$ is the sound speed and $v_A$ is the Alfv\'en speed). 
The plasma in the cavity is redirected along the curved shock towards the $z$-direction, where it forms a conical shock at the tip of the cavity, and a jet-like flow in the yz-plane is finally created \citep{ciardi2013astrophysics,albertazzi2014laboratory}.
Note that the tip of the shock envelope is actually not along the $y=0$ axis. There's a small bending towards the right, as is shown by the black velocity arrows. This is due to the initial non-axisymmetric momentum perturbation mentioned above.

The characterization of the shock envelope is detailed in the second column of Table \ref{tab:paras}. The collisionality (the ratio of the mean-free-path $\lambda_{mfp}$ to the shock envelope's length-scale $L_0$) of both electrons and ions is smaller than unity, confirming the validity of the fluid description \citep{ryutov1999similarity}. The Hall parameter (or, the magnetization) of electrons is $H_e = \lambda_{mfp,e}/r_{L,e} \sim 5$, indicating that the electrons are strongly magnetized; while ions are not ($H_i = \lambda_{mfp,i}/r_{L,i} \sim 0.03$). 

\begin{figure}[htp]
    \centering
    \includegraphics[width=0.3\textwidth]{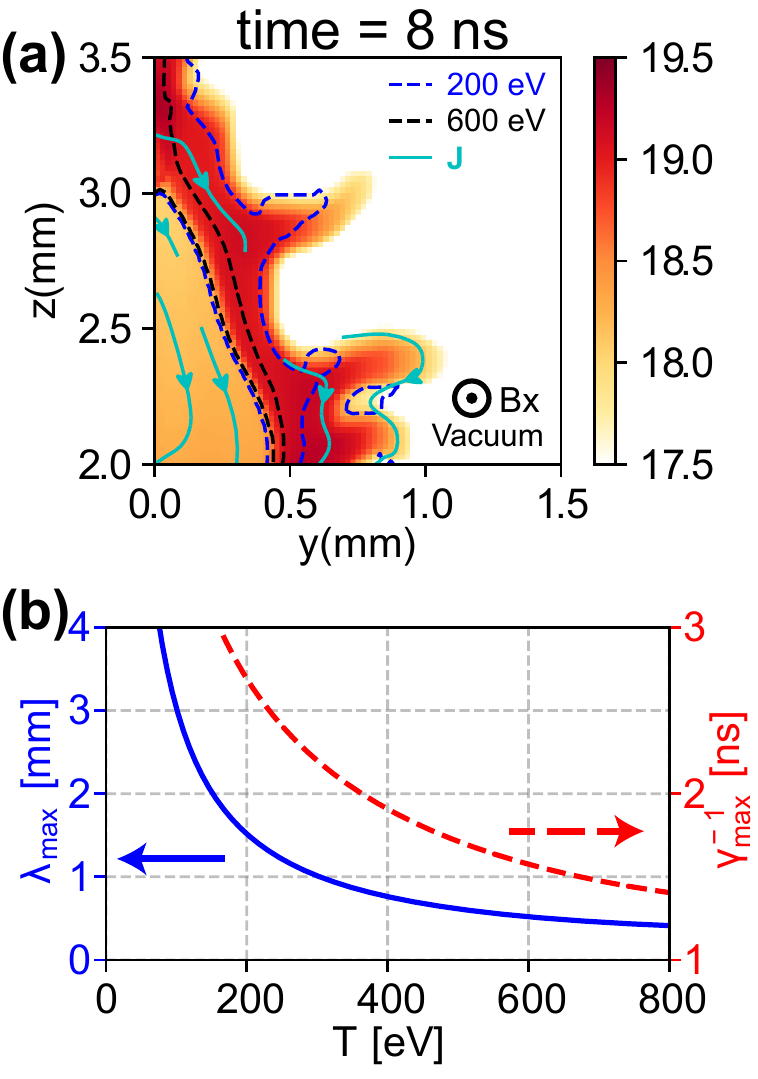}
    \caption{\textbf{MRTI growth during the slab formation.} (a) Zoom on the plasma/vacuum interface at $t = 8$ ns. The color map corresponds to $\log_{10}n_e$ in cm$^{-3}$. Cyan lines show the contours of the current density magnitude, whereas the dashed line contours show the ion temperature $T_i$. 
    (b) The temperature dependence of the fastest growing mode and the growth time for the MRTI.
    }
    \label{fig:MRTI}
\end{figure}

The MRTI requires an effective acceleration or deceleration at an interface between fluids of different mass density, and this basic condition is met at the interface between the plasma and vacuum when the expanding plasma is halted by the magnetic field lines and also when the cavity collapses\cite{khiar2017laboratory}.
As is seen in Fig.~\ref{fig:MRTI} (a), a zoom of the structure of the flow along the plasma/vacuum interface shows the growth of the protruding ``fingers'', which is one of the characteristic of the MRTI. 
This inevitably will be triggered if the effective acceleration $g_{eff}$ is anti-parallel to the density gradient (in the frame of the interface). 
Following the analytical estimate made in our former work \cite{khiar2019laser}, in Fig.~\ref{fig:MRTI} (b), the MRTI growth time and fastest growing mode are calculated with the parameters of our plasma condition at $t=8$ ns and at the interface, i.e. the mass density $\rho = 0.06$ kg/m$^{3}$, the Coulomb logarithm $\Lambda=9$, the atomic number $\langle A\rangle = 17.32$, the effective charge state $\langle Z\rangle = 8$, and the effective acceleration $g_{eff} \sim v_{\perp}^2 / \delta_{sl} \sim 3.3\times 10^{13}$ m/s$^2$, in which $\delta_{sl} \sim 300\  \mu$m is the width of the shock envelope and $v_{\perp} \sim 100$ km/s is the flow velocity. 
It is clear that over the temperature range of $200 - 800$ eV, as the temperature increases, the wavelength of the fastest growing mode flattens to a narrow band of $\lambda_{max} \sim 1$ mm, and for these modes the e-folding time is less than 3 ns, which is consistent with the simulations.

\begin{figure}[htp]
    \centering
    \includegraphics[width=0.48\textwidth]{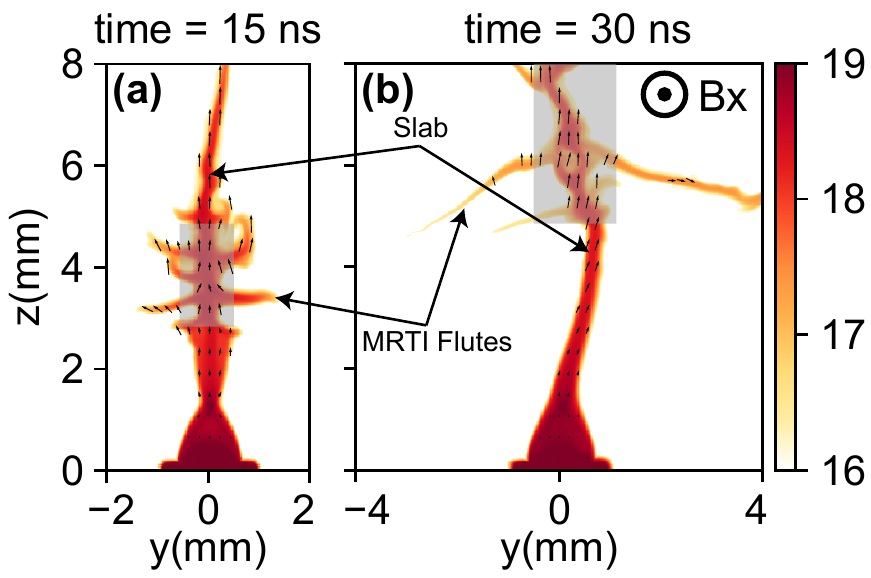}
    \caption{\textbf{Slab formation} (a) The collapse of the cavity at $t=15$ ns around $3 < z < 5$ mm (highlighted by the grey box). 
    (b) The fully grown flutes at $t=30$ ns reach the boundary of $y=4$ mm. 
    The collapsed region propagates along the $z$-direction, arriving at $5 < z < 8$ mm and the slab is formed.
    The color map corresponds to $\log_{10} \int n_e dx$ in cm$^{-2}$. 
    The black arrows are the velocity vectors.
    }
    \label{fig:flutes}
\end{figure}

The MRTI initially grows on the outer edges of the cavity but also propagates axially along with the flow. 
After the initial growth phase, Fig.~\ref{fig:flutes} (a) shows that at $t=15$ ns, the density cavity collapsed around $3 < z < 5$ mm, where the MRTI grows rapidly (highlighted by the grey box).
At $t=30$ ns, as is shown in Fig.~\ref{fig:flutes} (b), the collapsed region propagates along the $z$-direction, arriving at $5 < z < 8$ mm (also highlighted by the grey box). The MRTI flutes end up merging with each other, and elongate along the $y$-direction, reaching $y=4$ mm. 
In the meantime, their density is dropping, leaving the plasma plume density compressed around $y=0$ while propagating along the $z$-direction, thus forming a dense slab structure whose physical parameters are given in the third column of Table~\ref{tab:paras}.

\subsection{SLAB PROPAGATION}
\label{propagation}

We now describe the late-time slab propagation stage, focusing on the kink-like disruption see for example in Fig.~\ref{fig:3Ddemo} (c) and (d), as well as in Fig.~\ref{fig:2Ddemo}, which was clearly observed in recent experiments\cite{filippov2021enhanced}.

Comparing Fig.~\ref{fig:2Ddemo} (A3b) and (A4b), it is clear that the propagation of the MRTI-flutes (both in the $z$ and $y$ directions) leads to a decrease of their mass density, which eventually drops below the vacuum cut off employed in the simulations. Thus at late times, only the central portion of the slab, localized around $y=0$, is visible.
Additionally, we note that in the slab the electron and ion have equilibrated to equal temperatures.
Other parameters, e.g. the magnetic field, velocity, etc., are listed in Table~\ref{tab:paras}. 

\begin{figure}[htp]
    \centering
    \includegraphics[width=0.32\textwidth]{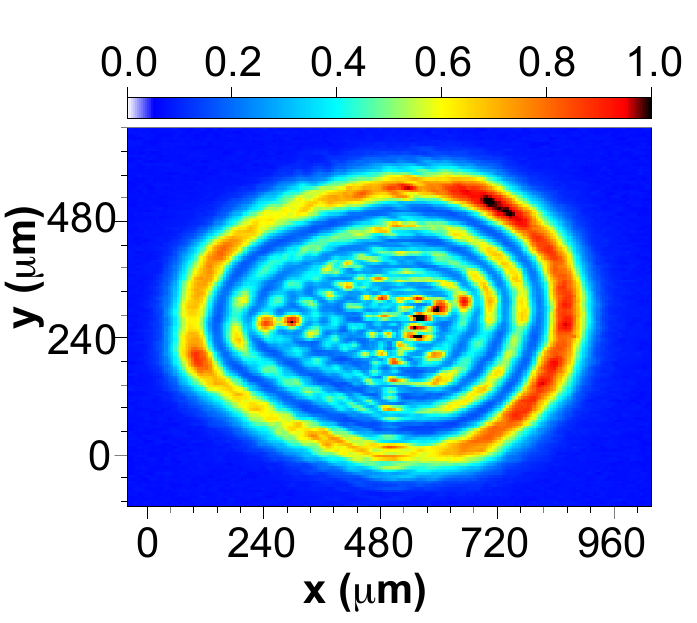}
    \caption{\textbf{Typical spatial pattern of the laser intensity deposition on target used in laboratory experiments.} The pattern is due to the laser is actually defocused on target, so that its intensity is not too large. The colormap represents the normalized intensity. 
    }
    \label{fig:Laser_spot}
\end{figure}

More interestingly, starting from Fig.~\ref{fig:2Ddemo} (3b), it is clear that a kink-like perturbation develops during the slab propagation. Such bending originates from the asymmetric perturbation introduced initially on the plasma momentum.
This is inspired by related experiments \citep{khiar2019laser}, where the typical spatial pattern of the laser intensity deposition on target is shown in Fig.~\ref{fig:Laser_spot}. Beside the relatively less important ring-like radial distributions, the intensity is not axis-symmetric with a clear left to right difference.

We now investigate the effects of an non-homogeneous laser-intensity deposition on the distortions of the slab. The purpose of the simulations is to demonstrate that the kinking of the slab can be simply reproduced by perturbing the initial velocity of the plasma plume, without the added complexity of introducing a non-homogeneous laser intensity profile. Although idealized, these initial conditions demonstrate that they are sufficient to perturb the oblique shock that is then responsible for redirecting the flow axially. In the simulation, the choice of the perturbation on the right being an order-of-magnitude higher than that on the left is to clearly demonstrate this effect within the simulation timescale.

\begin{figure}[htp]
    \centering
    \includegraphics[width=0.48\textwidth]{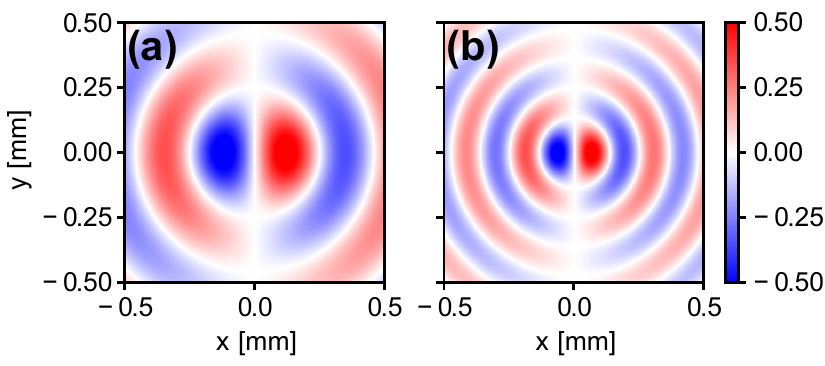}
    \caption{\textbf{Spatial distribution of the Bessel-like perturbation of the plasma velocity.} (a) $u_{1,1}$ with the mode of $m=1, n=1$. (b) $u_{1,2}$ with the mode of $m=1, n=2$. 
    Other parameters are: $r_0 = 0.5$ mm, $\alpha_{1,1} = 3.8317$, $\alpha_{1,2} = 7.0156$.
    The color map corresponds to the perturbation strength. 
    }
    \label{fig:Bessel_function}
\end{figure}

To mimic the ring-like radial intensity distribution, we introduce the following perturbation of the plasma velocity $u$, with a Bessel-like form:  
$$u_{m n}(r, \theta)= J_{m}\left(\frac{\alpha_{m n}}{r_0} r\right)\sin (m \theta)$$
where $r = (x^2 + y^2)^{1/2}$ is the distance to the circle of the laser spot on the target surface ($x$ and $y$ are the spatial coordinates), the angle $\theta = \arctan(y/x)$, $J_m$ is the Bessel function of the first kind, and $\alpha_{m n}$ is the n-th root of $J_m$, $r_0$ is the laser spot radius, as is illustrated in Fig.~\ref{fig:Bessel_function}.

\begin{figure}[htp]
    \centering
    \includegraphics[width=0.32\textwidth]{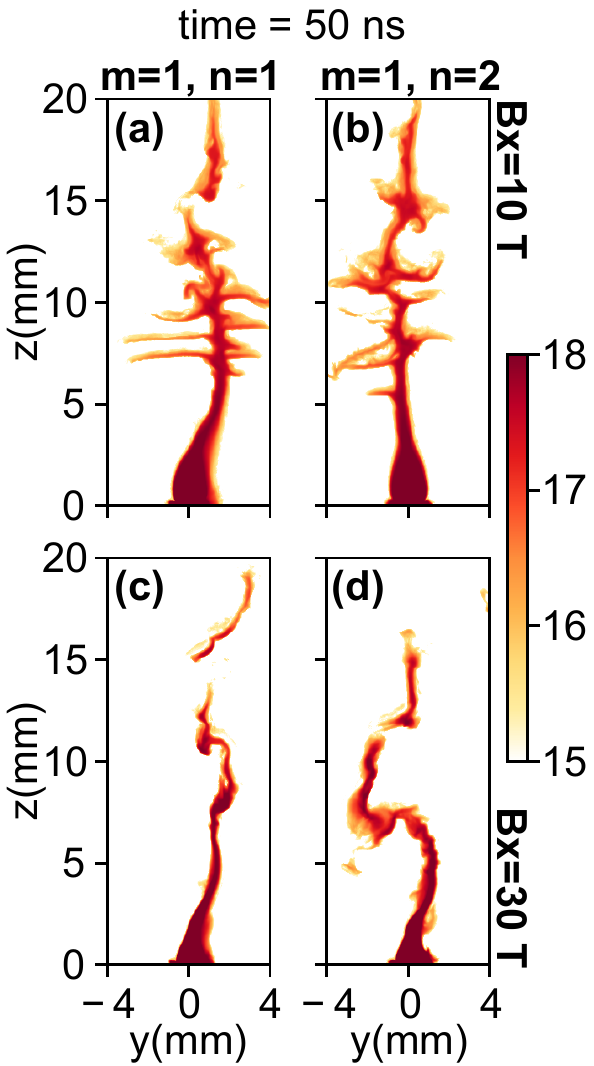}
    \caption{\textbf{Slab propagation with different modes of Bessel-like perturbation under different magnetic field strength at the end of the simulation at $t=50$ ns.} The colormap corresponds to $\log_{10} \int n_e dx$ in cm$^{-2}$, i.e., to the decimal logarithm of the electron number density integrated along the external magnetic field. 
    The first row is for $B_x = 10$ T, while the second is for $B_x = 30$ T.
    For the first column, the Bessel function has a mode of $m=1, n=1$, while for the second column, the mode is $m=1, n=2$.
    }
    \label{fig:Bessel_perturb}
\end{figure}

Here, we compare the results of the above two modes (i.e. $m=1, n=1$ and $m=1, n=2$) for different magnetic field strength (i.e. $B_x = 10$T and $B_x = 30$T) at $t=50$ ns.
Comparing the first row of Fig.~\ref{fig:Bessel_perturb}, with $B_x = 10$ T, the difference between the slab propagation with initial perturbation using Bessel function modes of $m=1, n=1$ and $m=1, n=2$ is not so obvious. They both show a weak slab bending along the z-direction, and there still exists quite a lot of flutes structures.
However, in the second row with $B_x = 30$ T, distinctive differences between those modes can be seen. Besides the bending of the slab, we also see the formation of the kink-like disruption, with patterns much stronger in the $m=1, n=2$ mode than that in the $m=1, n=1$ mode. The disruption leads to a shorter propagation distance for the $m=1, n=2$ mode than that of the $m=1, n=1$ mode. The kink-like disruption is clearly more severer with higher magnetic field strength. Note the the simulations shown here are different from the case shown in Fig.~\ref{fig:2Ddemo} (5b), in which the perturbation had no ring-like structure. This further indicates that the particular pattern of the laser spot inhomogeneity plays a crucial role in determining the pattern and the level of the kink-like disruptions. As mentioned above, this is also in accordance with recent experimental observations \citep{filippov2021enhanced}.

\begin{figure}[htp]
    \centering
    \includegraphics[width=0.45\textwidth]{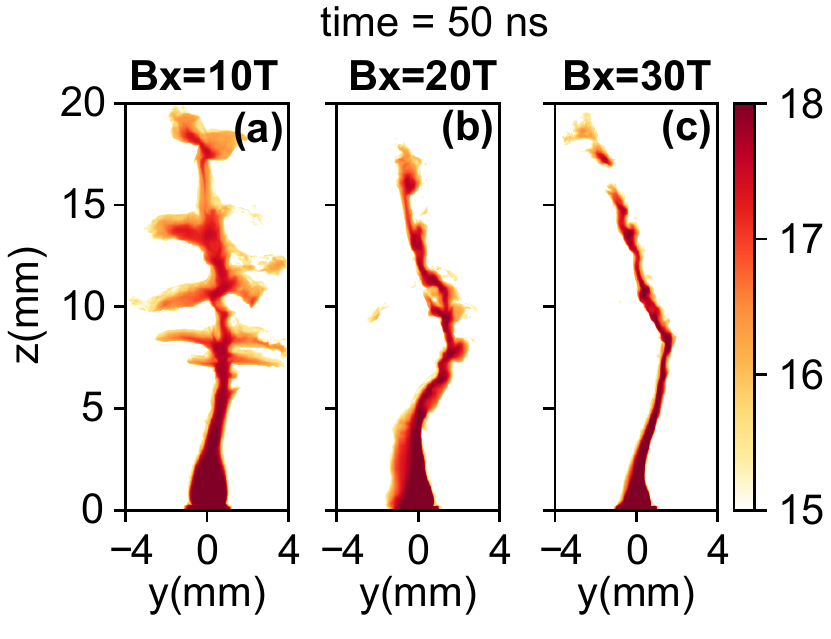}
    \caption{\textbf{Slab propagation with different initial magnetic field strength.} (a)-(c) $B_{x0} = 10/20/30$ T.
    The color map corresponds to $\log_{10} \int n_e dx$ in cm$^{-2}$. 
    }
    \label{fig:B-strength}
\end{figure}

To further investigate the effect of the magnetic field strength on the propagation of the plasma slab, we carried out a series of simulations with a magnetic field increasing from 10~T to 30~T and for fixed laser-plasma interaction conditions (i.e. the same as described in Sec.~\ref{SIMU}). The results are compared in Fig.~\ref{fig:B-strength}.
It is clear that as the magnetic field strength is increased the slab becomes thinner, its density increases, and the amplitude of the kink-like motions became more evident. In addition, as the MRTI is more suppressed with higher magnetic field, the flute structures in the $B_x = 10$ T case are much more obvious than those in the $B_x = 20$ T case, while they completely disappear in the $B_x = 30$ T case. Because as the magnetic field strength increased, the diamagnetic cavity collapses more quickly, leaving shorter time for the MRTI to grow. 

\section{CONCLUSIONS}
\label{CONC.}

In conclusion, the detailed characterization of the overall stability and dynamics of a laser-produced plasma across a magnetic field of tens of Tesla is investigated with large-scale 3D resistive-MHD simulations. 
It is found that the plasma is first collimated by the magnetic field into a slender slab, whose plasma-vacuum interface is dominated by the MRTI. 
Later, the flutes fade away, the plasma elongates forming a slab-like structure. 
We have shown that during its propagation, the slab suffers from kink-like disruptions because of the imbalance of initial perturbation, which originates from the inhomogeneity of the laser-intensity deposition. By mimicking the pattern of a typical laser-intensity profile from related experiment using a Bessel-like function, we find that the kink-like structures observed in the slab are closely linked to the initial velocity perturbations imposed and to the strength of the magnetic field.  


\section*{ACKNOWLEDGMENTS}
This work was supported by funding from the European Research Council (ERC) under the European Unions Horizon 2020 research and innovation program (Grant Agreement No. 787539). The research leading to these results is supported by Extreme Light Infrastructure Nuclear Physics (ELI-NP) Phase II, a project co-financed by the Romanian Government and European Union through the European Regional Development Fund, and by the project ELI-RO-2020-23 funded by IFA (Romania). This work was also granted access to the HPC resources of MesoPSL financed by the Region Ile de France and the project Equip @ Meso (reference ANR-10-EQPX-29-01) of the program Investissements d'Avenir supervised by the National Agency for Research.

\bibliographystyle{apsrev4-2-titles}
\bibliography{long}

\end{document}